# Long-Range Chiral Pairing enables Topological Superconductivity in Triangular Lattices without Spin-Orbit Coupling and Magnetic Field

Y. Z. Li[a], Y.Y. Lu[d✉], J. X. Zhong[c✉] and L. J. Meng[b✉]

[a]School of Physics and Electronic Science, Hunan Institute of Science and Technology, Yueyang 414006, People's Republic of China

[b]School of Physics and Optoelectronics, Xiangtan University, Xiangtan 411105, Hunan, People's Republic of China

[c]Center for Quantum Science and Technology, Department of Physics, Shanghai University, Shanghai 200444, People's Republic of China

[d]School of Computer and Information Engineering, Guizhou University of Commerce, Guizhou 550014, China

## Abstract

This paper demonstrates a pathway to topological superconductivity in monolayer triangular lattices through long-range pairing without requiring spin-orbit coupling and magnetic field, contrasting conventional frameworks reliant on superconductivity and spin-orbit coupling and time-reversal symmetry (TRS) breaking. Berry curvature analysis reveals spontaneous TRS-breaking-induced peaks or valleys under long-range pairing, signaling nontrivial topology superconducting state. Notably, the increase in the long-range pairing strength only changes the size of the energy band-gap, without triggering a topological phase transition. This characteristic is verified by calculating Berry curvature and topological edge states. In zigzag and armchair-edge ribbons of finite width, the topological edge states are regulated by the ribbon boundary symmetry and the interact range of long-range pairing. Under nearest-neighbor pairing, the topological edge states maintain particle-hole symmetry and matches the corresponding Chern number. However, next-nearest-neighbor and third-nearest-neighbor pairings break the particle-hole symmetry of the topological edge states in armchair-edge ribbon. This work proposes a mechanism for realizing topological superconductivity without relying on spin-orbit coupling and magnetic field, offering a theoretical foundation for simplifying the design of topological quantum devices.

## 1.Introduction

Topological superconductors exhibit bulk band-gap induced by superconducting pairing potentials and topological edge states hosting non-Abelian statistics Majorana fermions, which show promising potential in the field of topological quantum computing[1-3]. The most straightforward way to implement topological superconductors is to search for odd parity p-wave superconductors, which are extremely rare in nature. Researchers have subsequently turned to the study of artificial topological superconductors and proposed several schemes to achieve topological superconductivity[4], such as the proximity effect[5], chemical doping[6-8], and external field modulation[9], etc.. However, these external control methods generally require precise control of external conditions, and their effects may be unstable. And previous study also shows local (say, electric or magnetic) fields do not manipulate the quantum information[10].

---

✉ The corresponding author Email: xiaoguai@gzsxy361.wecom.work
✉ The corresponding author Email: jxzhong@xtu.edu.cn
✉ The corresponding author Email: ljmeng@xtu.edu.cn

Theoretical research indicates that Majorana zero mode-dependent odd-parity pairing (such as p-wave pairing) often involves electron interactions between different lattice sites (i.e., long-range pairing)[11]. Recent reports have shown that, under long-range pairing superconducting states, time-reversal symmetry (TRS) can be spontaneously broken, inducing a topological superconducting (TSC) state under zero magnetic field, which significantly reduces the experimental requirements[12-17]. Of particular note is that long-range interactions can not only serve as an alternative path to topological superconductivity but can also actively enhance the topological order. As demonstrated by Viyuela et al. in a two-dimensional p-wave superconductor with long-range hopping and pairing amplitudes, long-range couplings can significantly enlarge the regime of a topological chiral phase in the parameter space[18]. This "enhancement" effect dramatically reduces the experimental demand for precise tuning of the chemical potential, providing a broader platform for realizing and observing topological superconducting states. Monolayer $MoS_2$, considering the nearest-neighbor (NN) spin-singlet pairing potential, TSC phases with non-zero Chern number (CN) and spontaneous TRS breaking is realized[14]. The graphene model with NN pairing presents TSC state with CNs of 1 and 2 under zero magnetic field[15]. The two-dimensional $D_{4h}$ point symmetric square lattice achieves a mixed singlet superconducting pairing function that combines on-site and long-range pairing, exhibiting topologically nontrivial high CNs and Majorana zero modes located outside high symmetry points[16]. The checkerboard model with next-nearest-neighbor (NNN) pairing displays multiple TSC states with CN as high as 4 when the net magnetic field is zero[17]. However, in certain cases, long-range pairing superconductors have numerous non-topological energy bands near the Fermi level ($E_F$). An excess of non-topological energy bands can easily obscure the topological electronic behavior. Moreover, topological edge states with some CNs are not always robust, and CNs and topological edge states does not always adhere to the conventional bulk-boundary correspondence principle[1]. The underlying physical mechanism behind this mismatch between CN and topological edge state count remains unclear and requires comprehensive and in-depth research.

In this paper, we obtain long-range pairing functions (including NN, NNN, and third-nearest-neighbor (TNN) pairings) with higher-dimensional irreducible lattice representations $E_2$ for the triangular lattice with $C_{6v}$ point group symmetry by using the projection operator method[19]. The singlet long-range pairing function corresponds to d-wave symmetry and naturally suggests possible chiral superconductivity phases[20]. Potential applications of anisotropic d + id superconducting pairings have been proposed in materials such as water-intercalated sodium cobaltates, bilayer silicene, the epitaxial bilayer films of bismuth and nickel.[21-23]. To facilitate the experimental regulation, this paper investigates systematically TSC state by considering chiral d+id' pairing under zero magnetic field based on a Bogoliubov-de Gennes (BdG) Hamiltonian. We initially employ efficient method[24] to study CN as a function of chemical potential and long-range pairing strength. Subsequently, we calculate the topological edge states of zigzag and armchair ribbons to confirm TSC state of CNs phase diagram. To comprehensive study of the topological edge state of the system, we further investigated the probability distributions $|\psi(n)|^2$ near the $E_F$.

**2. Model and Method**

The model under consideration is a triangular lattice with point group $C_{6v}$. The paper primarily interested in studying the chiral superconducting states d+id′, which can be modeled as pairing on NN and NNN and TNN neighbors. The pairings are initially incorporated into our model, so the tight-binding Hamiltonian defined on a triangular lattice becomes[25]

$$H = H_t + H_{sc} \tag{1}$$

$$H_t = -\sum_{i,j,\sigma} t_{ij} c_{i\sigma}^\dagger c_{j\sigma} - \mu \sum_{i,\sigma} c_{i\sigma}^\dagger c_{i\sigma} \tag{2}$$

$$H_{sc} = \sum_{\langle i,j \rangle} \Delta_{ij} \left( c_{i\uparrow}^\dagger c_{j\downarrow}^\dagger + c_{i\downarrow} c_{i\uparrow} \right) \tag{3}$$

where the hopping term $H_t$ contains the hopping amplitudes $t_{ij}$ and the chemical potential $\mu$. The last term describes long-range pairing, and the $<i, j>$ describe the $i$ and $j$ neighboring pairings (including NN, NNN or TNN).

Previous study introduces long-range pairing potential based on the projection operator approach[26]. The trial wave functions of three types of neighbors (NN, NNN, TNN in Fig. 1(a)) are

$$\begin{aligned} \phi_1^{NN} &= \delta_{i,i+x}, \quad \phi_2^{NN} = \delta_{i,\left(i+\frac{x}{2},i+\frac{\sqrt{3}y}{2}\right)} \\ \phi_1^{NNN} &= \delta_{i,i+\sqrt{3}y}, \quad \phi_2^{NNN} = \delta_{i,\left(i-\frac{3x}{2},i+\frac{\sqrt{3}y}{2}\right)} \\ \phi_1^{TNN} &= \delta_{i,i+2x}, \quad \phi_2^{TNN} = \delta_{i,\left(i+x,i+\sqrt{3}y\right)} \end{aligned} \tag{4}$$

We make use of a fundamental projection theorem stating that the operator $p(E_2) = \sum_g \chi_i^*(g) g$ projects out the contribution which transforms in the irreducible representation $E_2$. Here, the sum runs over all point-group operations g with the corresponding complex-conjugate characters $\chi_i^*(g)$. we then apply the projection operator $p(E_2)$ to trial wave functions Eq. (4) and obtain the following NN and NNN and TNN basis functions for the trivial representation $E_2$:

$$p(E_2)\phi_1^{NN} = \chi_i^*(E)\delta_{i,i+x} + \chi_i^*(C_2)\delta_{i,i-x} + \chi_i^*(C_3)\delta_{i,\left(i-\frac{x}{2},i+\frac{\sqrt{3}y}{2}\right)}$$
$$+ \chi_i^*(C_3^{-1})\delta_{i,\left(i-\frac{x}{2},i-\frac{\sqrt{3}y}{2}\right)} + \chi_i^*(C_6)\delta_{i,\left(i+\frac{x}{2},i+\frac{\sqrt{3}y}{2}\right)} + \chi_i^*(C_6^{-1})\delta_{i,\left(i+\frac{x}{2},i-\frac{\sqrt{3}y}{2}\right)}$$
$$p(E_2)\phi_2^{NN} = \chi_i^*(E)\delta_{i,\left(i+\frac{x}{2},i+\frac{\sqrt{3}y}{2}\right)} + \chi_i^*(C_2)\delta_{i,\left(i-\frac{x}{2},i-\frac{\sqrt{3}y}{2}\right)} + \chi_i^*(C_3)\delta_{i,i-x}$$
$$+ \chi_i^*(C_3^{-1})\delta_{i,\left(i+\frac{x}{2},i-\frac{\sqrt{3}y}{2}\right)} + \chi_i^*(C_6)\delta_{i,\left(i-\frac{x}{2},i+\frac{\sqrt{3}y}{2}\right)} + \chi_i^*(C_6^{-1})\delta_{i,i+x}$$
(5)

$$p(E_2)\phi_1^{NNN} = \chi_i^*(E)\delta_{i,i+\sqrt{3}y} + \chi_i^*(C_2)\delta_{i,i-\sqrt{3}y} + \chi_i^*(C_3)\delta_{i,\left(i-\frac{3}{2}x,i-\frac{\sqrt{3}}{2}y\right)}$$
$$+ \chi_i^*(C_3^{-1})\delta_{i,\left(i+\frac{3}{2}x,i-\frac{\sqrt{3}}{2}y\right)} + \chi_i^*(C_6)\delta_{i,\left(i-\frac{3}{2}x,i+\frac{\sqrt{3}}{2}y\right)} + \chi_i^*(C_6^{-1})\delta_{i,\left(i+\frac{3}{2}x,i-\frac{\sqrt{3}}{2}y\right)}$$
$$p(E_2)\phi_2^{NNN} = \chi_i^*(E)\delta_{i,\left(i+\frac{3}{2}x,i+\frac{\sqrt{3}}{2}y\right)} + \chi_i^*(C_2)\delta_{i,\left(i-\frac{3}{2}x,i-\frac{\sqrt{3}}{2}y\right)i-\sqrt{3}y} + \chi_i^*(C_3)\delta_{i,\left(i-\frac{3}{2}x,i+\frac{\sqrt{3}}{2}y\right)}$$
$$+ \chi_i^*(C_3^{-1})\delta_{i,i-\sqrt{3}y} + \chi_i^*(C_6)\delta_{i,i+\sqrt{3}y} + \chi_i^*(C_6^{-1})\delta_{i,\left(i+\frac{3}{2}x,i-\frac{\sqrt{3}}{2}y\right)}$$
(6)

$$p(E_2)\phi_1^{TNN} = \chi_i^*(E)\delta_{i,i+2x} + \chi_i^*(C_2)\delta_{i,i-2x} + \chi_i^*(C_3)\delta_{i,(i-x,i+\sqrt{3}y)}$$
$$+ \chi_i^*(C_3^{-1})\delta_{i,(i-x,i-\sqrt{3}y)} + \chi_i^*(C_6)\delta_{i,(i+x,i+\sqrt{3}y)} + \chi_i^*(C_6^{-1})\delta_{i,(i+x,i-\sqrt{3}y)}$$
$$p(E_2)\phi_2^{TNN} = \chi_i^*(E)\delta_{i,(i+x,i+\sqrt{3}y)} + \chi_i^*(C_2)\delta_{i,(i-x,i-\sqrt{3}y)} + \chi_i^*(C_3)\delta_{i,i-2x}$$
$$+ \chi_i^*(C_3^{-1})\delta_{i,(i+x,i-\sqrt{3}y)} + \chi_i^*(C_6)\delta_{i,(i-x,i+\sqrt{3}y)} + \chi_i^*(C_6^{-1})\delta_{i,i+2x}$$
(7)

By transforming Eqs. (5)-(7) into *k*-space, we obtain the following long-range pairing functions ($\Delta_{E2NN}$, $\Delta_{E2NNN}$, $\Delta_{E2TNN}$) Eqs. (8)-(11) for the two-dimensional representation $E_2$ with strength parameters. In real materials, pairing channels at different distances often coexist and interact with each other. As observed in the honeycomb lattice, the NN and NNN d+id-wave pairings can coexist and mutually promote each other[27]. Similarly, in the triangular lattice, NN, NNN, and TNN pairings all belong to the $E_2$ irreducible representation and thus can naturally mix. We adopt spherical coordinate parameterization, where the relative strength of the long-range pairing function is controlled by the azimuthal angle ($\theta$, $\varphi$) (as shown in Fig.1(b)).

$$\Delta_{E2NN}^{d_{x^2-y^2}}(\boldsymbol{k}) = 2\cos(k_x) - \cos\left(k_x/2 + \sqrt{3}k_y/2\right) - \cos\left(k_x/2 - \sqrt{3}k_y/2\right)$$
$$\Delta_{E2NN}^{d_{xy}}(\boldsymbol{k}) = -\cos(k_x) + 2\cos\left(k_x/2 + \sqrt{3}k_y/2\right) - \cos\left(k_x/2 - \sqrt{3}k_y/2\right)$$
(8)

$$\Delta_{E2NNN}^{d_{x^2-y^2}}(\boldsymbol{k}) = 2\cos(\sqrt{3}k_y) - \cos\left(3k_x/2 + \sqrt{3}k_y/2\right) - \cos\left(-3k_x/2 + \sqrt{3}k_y/2\right)$$
$$\Delta_{E2NNN}^{d_{xy}}(\boldsymbol{k}) = -\cos(\sqrt{3}k_y) + 2\cos\left(3k_x/2 + \sqrt{3}k_y/2\right) - \cos\left(-3k_x/2 + \sqrt{3}k_y/2\right)$$
(9)

$$\Delta_{E2TNN}^{d_{x^2-y^2}}(\boldsymbol{k}) = 2\cos(2k_x) - \cos\left(k_x + \sqrt{3}k_y\right) - \cos\left(k_x - \sqrt{3}k_y\right)$$
$$\Delta_{E2TNN}^{d_{xy}}(\boldsymbol{k}) = -\cos(2k_x) + 2\cos\left(k_x + \sqrt{3}k_y\right) - \cos\left(k_x - \sqrt{3}k_y\right)$$
(10)

$$E_{SC}(\boldsymbol{k}) = s_{total} \sin\theta \cos\varphi \left( \Delta_{E2NN}^{d_{x^2-y^2}}(\boldsymbol{k}) + i\Delta_{E2NN}^{d_{xy}}(\boldsymbol{k}) \right)$$

$$+ s_{total} \sin\theta \sin\varphi \left( \Delta_{E2NNN}^{d_{x^2-y^2}}(\boldsymbol{k}) + i\Delta_{E2NNN}^{d_{xy}}(\boldsymbol{k}) \right) \quad (11)$$

$$+ s_{total} \cos\theta \left( \Delta_{E2TNN}^{d_{x^2-y^2}}(\boldsymbol{k}) + i\Delta_{E2TNN}^{d_{xy}}(\boldsymbol{k}) \right)$$

here, $s_{total} = \sqrt{(s_{total}\sin\theta\cos\varphi)^2 + (s_{total}\sin\theta\sin\varphi)^2 + (s_{total}\cos\theta)^2}$ represents the total pairing strength. The polar angle $\theta \in [0, \pi/2]$ controls the ratio of TNN to in-plane pairing (NN and NNN), while the azimuthal angle $\varphi \in [0, \pi/2]$ regulates the relative weight of NN and NNN pairings. We use ($s_{total}$, $\theta$, $\varphi$) to denote the long-range pairing. Especially, $\theta=\pi/2$ and $\varphi=0$ correspond to pure NN pairing dominance, $\theta=\pi/2$ and $\varphi=\pi/2$ correspond to pure NNN pairing dominance, $\theta=0$ and $\varphi=0$ correspond to pure TNN pairing dominance, while intermediate angles represent various mixed configurations.

Then, the BdG Hamiltonian of triangular lattice in the Nambu basis $\psi=(c_{k\uparrow}, c_{k\downarrow}, c^\dagger_{-k\uparrow}, c^\dagger_{-k\downarrow})^T$, incorporating hopping term $E_t(\boldsymbol{k})$ and long-range pairing term $E_{sc}(\boldsymbol{k})$, is obtained as follows:

$$H(\boldsymbol{k}) = \begin{bmatrix} E_t(\boldsymbol{k}) - \mu & 0 & 0 & E_{SC}(\boldsymbol{k}) \\ 0 & E_t(\boldsymbol{k}) - \mu & -E_{SC}(\boldsymbol{k}) & 0 \\ 0 & -E_{SC}^\dagger(\boldsymbol{k}) & -E_t(-\boldsymbol{k}) + \mu & 0 \\ E_{SC}^\dagger(\boldsymbol{k}) & 0 & 0 & -E_t(-\boldsymbol{k}) + \mu \end{bmatrix} \quad (12)$$

where the hopping term $E_t$ includes the NN, NNN and TNN with parameters $t_1$, $t_2$ and $t_3$. In numeric, $t=t_1=1.0$ is set as unity and $t_2 = 0.1t_1$ and $t_3 = 0.01t_1$. $E_{sc}(\boldsymbol{k})$ is the long-range pairing potential, including NN, NNN and TNN pairings.

To study topological properties of TSC phase, we initially employ efficient method[24] to compute CN as a function of chemical potential and long-range pairing strength. The CN can be given by

$$CN = \frac{1}{2\pi i} \sum_{k_\ell} \ln\left[ U_1(k_\ell) U_2(k_\ell + \hat{1}) U_1(k_\ell + \hat{2})^{-1} U_2(k_\ell)^{-1} \right] \quad (13)$$

$$U_\nu(k_\ell) = \frac{\langle n(k_\ell) | n(k_\ell + \hat{\nu}) \rangle}{|\langle n(k_\ell) | n(k_\ell + \hat{\nu}) \rangle|} \quad (14)$$

where $\hat{\nu}$ is a vector in the direction, $k_\ell$ denotes lattice points on the discrete Brillouin zone. $|n(k_\ell)\rangle$ is a wave function of the $n$th Bloch band.

Fig. 1(c)-(e) show that as the pairing distance increases, the phase distribution of pairing function becomes more diverse and complex, and the positions of sudden phase jumps increase. In the absence of a long-range pairing potential, this studied model displays a trivial insulator as plotted in Fig.1 (f)-(g).

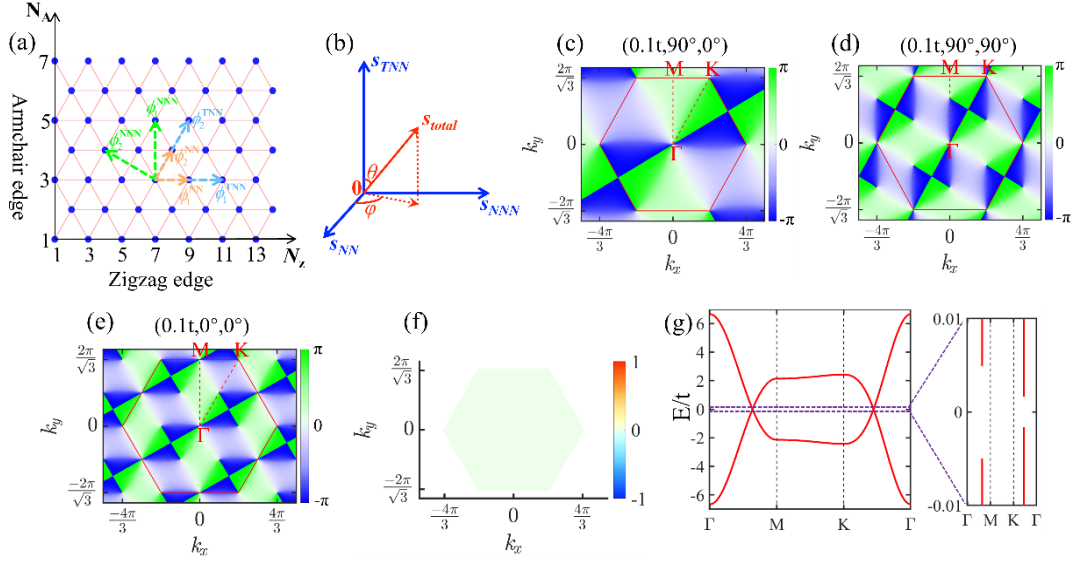

Fig.1 (a) The trigonal lattice and zigzag/armchair ribbon with width $N_Z/N_A$. The orange/green/blue arrow denotes the initial projection vector of NN/NNN/TNN pairing. (b) the total pairing strength $s_{total}$ in azimuth angle $(\theta, \varphi)$ in the spherical coordinates. The phase distribution of momentum dependence of the long-range pairing function for (c) NN, (d) NNN, and (e) TNN pairing with strength of 0.1t. The color represents the phase angle $\theta(\boldsymbol{k}) = arctan\left(\Delta^{d_{xy}}(\boldsymbol{k}) / \Delta^{d_{x^2-y^2}}(\boldsymbol{k})\right)$ of the chiral pairing potential $\Delta(\boldsymbol{k}) = \Delta^{d_{x^2-y^2}}(\boldsymbol{k}) + i\Delta^{d_{xy}}(\boldsymbol{k})$. The solid red hexagon represents the first Brillouin zone. (f) The sum over the Berry curvature of the valence bands of the BdG Hamiltonian in the first Brillouin zone without a long-range pairing potential. (d) The bulk bands in the absence of a long-range pairing potential.

## 3.Results and Discussion

The triangular lattice exhibits TSC states considering long-range pairing even in the absence of spin-orbit coupling (Fig.2(a)-(c)), providing a less restrictive new pathway compared to current requirements for realizing topological superconductivity, which include superconductivity, spin-orbit coupling, and TRS breaking[9, 10, 28]. Fig. 2(a)-(c) demonstrate that as the $\mu$ changes, the band-gap undergoes closure and reopening, accompanied by variations in the CN. This is consistent with previous results that the closing and reopening of the band-gap drive topological phase transitions[29]. Under the NN pairing, the studied system shows TSC state with a CN of -4 in the $\mu$ range of [-6.6, 2.4]t. While considering the NNN pairing and TNN pairing (Fig. 2(b)-(c)), nonzero CNs of 4, -2, -8 and -4, 8 emerge as the $\mu$ increases respectively. It is worth noting that mixed pairings do not introduce new topological CNs beyond those already observed in the individual channels. This indicates that there is no cooperative effect among the NN, NNN, and TNN pairings (i.e., the emergence of new CN phases); instead, their relationship is competitive. The topological phase of TNN has the largest phase region (Fig. 2(d)-(f)), indicating that the TNN paired components dominate in the competition, which also proves that long-range pairing can promote topological superconductivity[18].

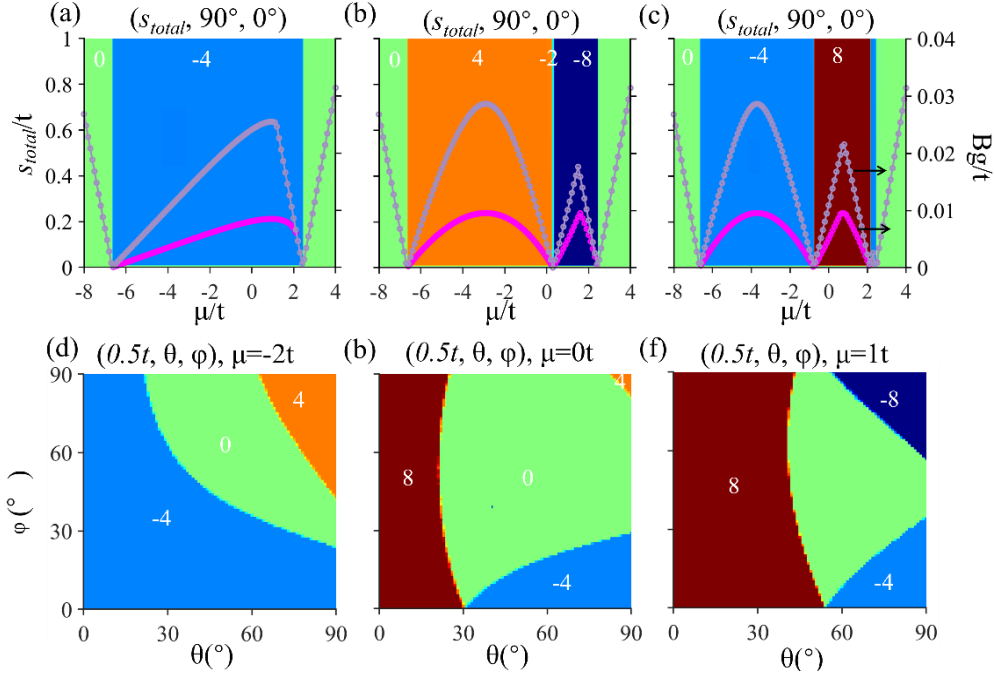

Fig.2 The TSC phase diagrams as a function of $\mu$ and $s_{total}$ for (a) NN, (b) NNN and (c) TNN pairings. Magenta and dark purple circles represent bulk band-gap with the pairing strength of 0.3t and 0.9t in (a)-(c), respectively. (d)-(e) The TSC phase diagrams as a function of azimuth angle ($\theta$, $\varphi$) for mixed pairing.

Berry curvature is determined by electronic band structures, where band inversion may induce sharp peaks or valleys in Berry curvature near inversion momentum points [30]. When considering long-range pairing, TRS is broken, leading to distinct peaks or valleys in the distribution of Berry curvature, suggesting the emergence of topological characteristic in the system (Fig. 3 (e)-(j)). When considering NN pairing, the band-gap along Γ-K is smaller than that along Γ-M, and band inversion is more likely to occur at the momentum path with a relatively small band-gap (Fig. 3(a)). The Berry curvature displays peaks along the Γ-K high-symmetry line, indicating that band inversion occurs on this path (Fig. 3(e)). In the superconducting states with NNN pairing and TNN pairing, both the band structures and Berry curvature distributions show that band inversion occurs along the Γ-M and Γ-K high-symmetry lines, respectively, as illustrated in Fig. 3(b, c, f, g). Interestingly, as shown in Fig. 2, the CN does not vary with increasing long-range pairing strength, which differs from previous results [12]. This is attributed to the fact that long-range pairing only alters the magnitude of band-gap and modifies slightly band shapes without inducing band-gap closure and reopening, thereby failing to drive a topological phase transition. Fig. 3(h)-(j) further demonstrate that as the superconducting pairing potential increases, the Berry curvature distribution does not exhibit abrupt changes, leaving the topological properties of the bands unaltered. For the mixed pairing ($s_{total}$, 20°, 10°) and ($s_{total}$, 60°, 10°), since the NN pairing and TNN pairing account for a large proportion, band inversion in the energy bands and Berry curvature are more likely to occur on the Γ-K path with a smaller band-gap (Fig.3(d,k,l)). This is the same as the results of pure NN pairing and TNN pairing. This also implies that no new topological CNs emerge in the system under mixed pairing, which is consistent with the aforementioned research results (as shown in Fig. 2(d)-(f)).

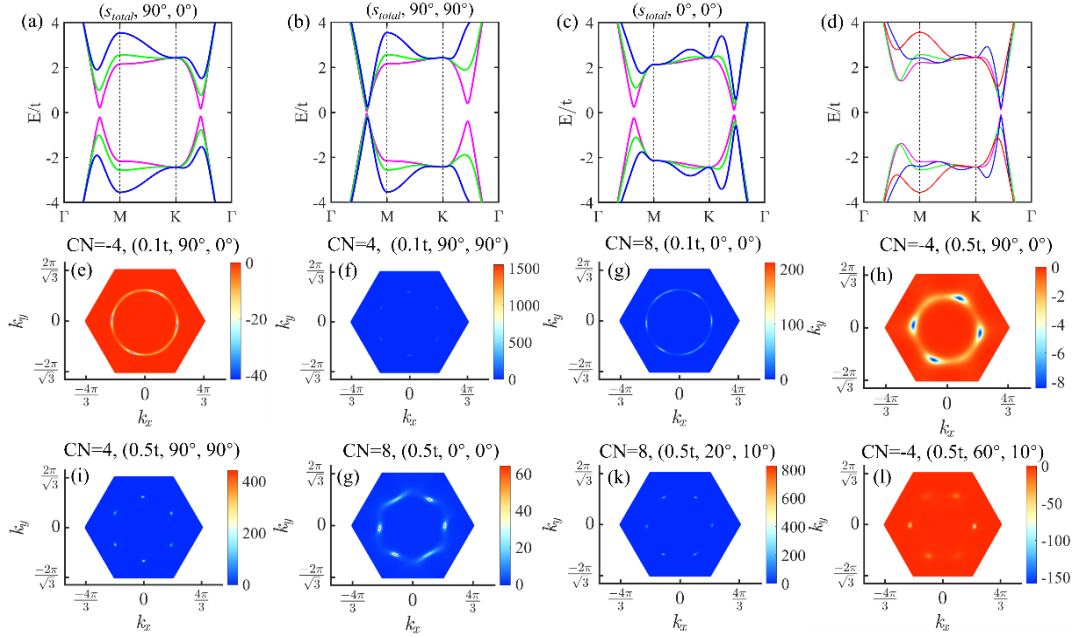

Fig.3 The bulk bands along the high-symmetry lines in the first Brillouin zone when considering different pairing potentials for (a) NN, (b) NNN, and (c) TNN pairing. The magenta, green, and blue lines correspond to pairing strengths of 0.1t, 0.5t, and 1.0t, respectively. (d) The bulk bands when considering the mixed long-range pairing. The magenta and blue lines represent the mixed pairing (0.5t, 20°, 10°) and (1.0t, 20°, 10°). The green and red lines represent the mixed pairing (0.5t, 60°, 10°) and (1.0t, 60°, 10°). The sum over the Berry curvature of the valence bands of the BdG Hamiltonian in the first Brillouin zone for (e)(h) NN and (f)(i) NNN and (g)(j) TNN and mixed (k)(l) pairing with $\mu = 0.0$.

To further study the topological property of the TSC state, we construct a tight-binding model for an infinitely long strip of triangular lattice with finite width 600 along zigzag and armchair edges (Fig. 1(a)). The two energy bands near the $E_F$ are doubly degenerate due to spin degeneracy as shown in the Fig. 4 and Fig. 5. The shape of the topological edge state depends on the symmetry of the ribbon, and thus exhibits different behaviors at different boundaries [31]. In the above analysis, band inversion occurs along the Γ-K and Γ-M (Fig. 3), which implies the location of the topological edge states. Long-range pairing is equivalent to adding "long-range hopping" in real-space, which will affect the band dispersion at the edges. The phase distribution of the NN pairing function in the first Brillouin zone is relatively smooth (Fig. 1(c)). In contrast, the phase distributions of the NNN and TNN pairing functions have more abrupt change positions (Fig. 1(d)-(e)), and the distribution of the Berry curvature is relatively scattered (Fig. 3(e, f, h, i)), which means that the topological edge state will also be relatively disordered and thus interact with each other. As shown in Fig. 4, the topological edge state of the zigzag strip exhibits obvious particle-hole symmetry, and the number of topological edge state corresponds to the CN. However, the particle-hole symmetry shows different behaviors in the topological edge state of the armchair ribbon. Under the NN pairing, the particle-hole symmetry of the armchair ribbon is preserved, and the system still has topological edge state corresponding to the CN. Under the NNN and TNN pairings, the particle-hole symmetry of topological edge state of the armchair ribbon is broken, and the topological edge state do not intersect at the $E_F$. As the long-range pairing strength increases, the topological edge band-gap progressively widens while leaving the system's topological properties unchanged. These results agree with the analytical results from Fig. 2 and Fig. 3, confirming that enhanced long-range pairing modifies band-gap magnitudes without altering the system's topological characteristics. In the case of mixed pairings ($s_{total}$, 20°, 10°) and ($s_{total}$, 60°, 10°), the topological edge states exhibit the same features as those induced by pure long-range pairings (Fig.4(g)-(h)). For the zigzag ribbon, the topological edge states remain robust and faithfully reflect the corresponding CN. In contrast, for the armchair ribbon, the breaking of particle-hole symmetry leads to asymmetric topological edge states dispersions that do not intersect at the $E_F$ (Fig.5(g)-(h)). The topological edge states do not display new types of crossings or additional degeneracies beyond those observed in individual channel cases. This behavior is consistent with the competitive relation revealed in the bulk phase diagrams (Fig. 2).

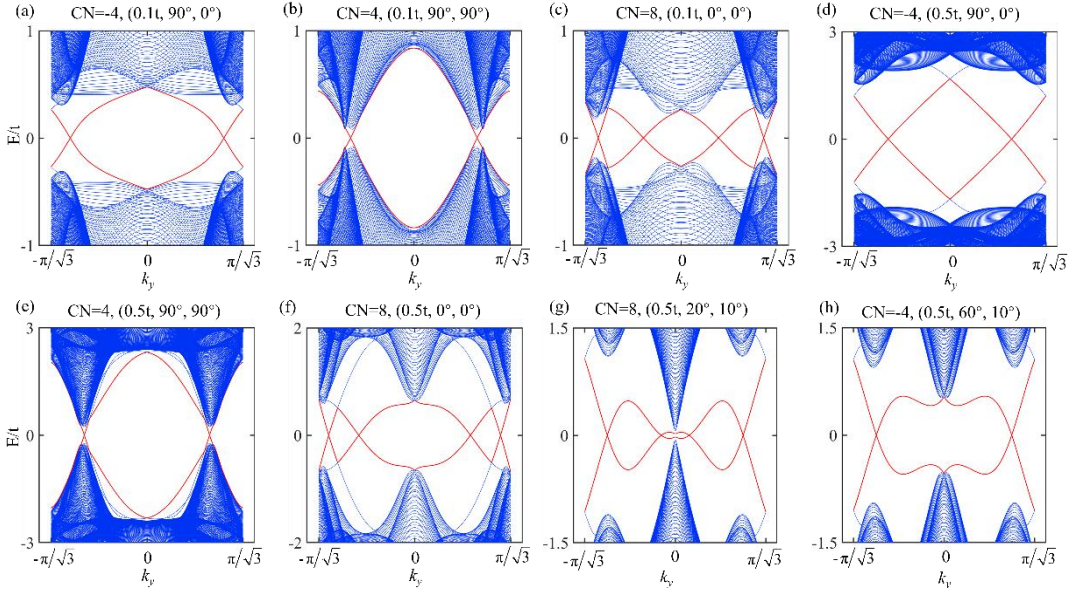

Fig.4 Band structures projected along the $k_y$ direction for zigzag ribbon with the NN (a)(d), NNN (b)(e) and TNN (c)(f) and mixed (g)(h) d+id′-wave pairings at $\mu$=0.0.

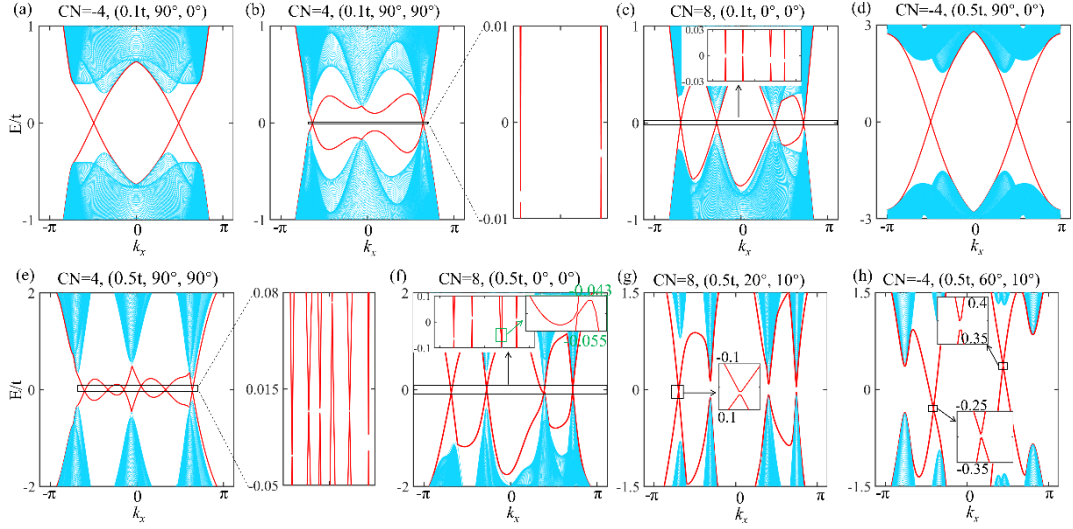

Fig.5 Band structures projected along the $k_x$ direction for armchair ribbon with the NN (a)(d), NNN (b)(e) and TNN (c)(f) and mixed (g)(h) d+id′-wave pairings at $\mu=0.0$.

To comprehensive study of the topological edge state of the system, we calculate the $|\psi(n)|^2$ of zigzag (Figs. 6(a)-(f)) and armchair ribbons (Figs. 6(g)-(l)), which display topological superconductivity in real space near the $E_F$. Under different pairing ranges and strengths, the real space $|\psi(n)|^2$ in zigzag and armchair ribbons exhibit clear edge-localized features. This indicates that the Majorana zero modes are well-separated and located at boundaries, consistent with previous results[17]. As the pairing function strength increases, the bulk superconducting gap of the system becomes larger, as shown in Fig. 3(a)-(c). A larger bulk band-gap can more effectively isolate the topologically protected edge states from the bulk states (see in Fig.4-Fig.5), helping to suppress the propagation of bulk states, making the topological edge states more localized, and thereby improving the purity and stability of the topological edge states. The enhancement of the localized characteristic of topological edge states also implies that the studied system exhibits stronger topological properties. This demonstrates that the effect of the pairing strength on both topological edge states and $|\psi(n)|^2$ is coherent and mutually consistent.

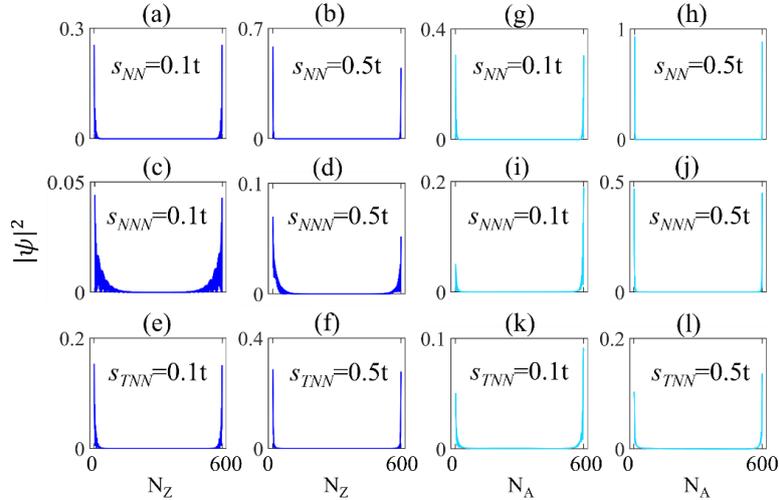

Fig.6 The $|\psi(n)|^2$ of (a)-(f) zigzag ribbon and (g)-(l) armchair ribbon near the $E_F$ in the real space when considering different pairing potentials with at $\mu=0.0$.

## 4. Conclusion

In conclusion, our study systematically investigates TSC states in a triangular lattice with long-range superconducting pairing, including NN, NNN, and TNN pairing. The spontaneous breaking of TRS under long-range pairing results in extreme distributions of Berry curvature and non-zero CNs, confirming the coexistence of topological and superconducting properties. Variations in $\mu$ induces band-gap closure and reopening accompanied by CN changes, consistent with bulk-boundary correspondence. Notably, the CN remains unchanged with increasing long-range pairing strength, indicating that long-range pairing modifies the band-gap size without altering the topological nature of the bands, thereby confirming the robustness of long-range superconducting states. The study of topological edge state show that the NN pairing preserves the particle-hole symmetry of the topological edge state along the $k_x$ and $k_y$ directions and corresponds to CN, while the particle-hole symmetry of the topological edge state along the $k_x$ direction for the NNN and TNN pairings is broken. This indicates that the boundary symmetry and the interact range of long-range pairing functions play a decisive role in the distribution of topological electronic states. The analysis of the $|\psi(n)|^2$ near the $E_F$ further indicates that as the pairing potential strength increases, the energy range of topological edge state become more wider, resulting in enhanced edge localization of the $|\psi(n)|^2$. Our study proposes a TSC state realization mechanism that does not rely on spin-orbit coupling and magnetic field, providing a feasible strategy for the experimental design of topological superconductors, and reducing the strict requirements for material constraints.

## Acknowledgements

This work is supported by Education Department of Hunan Province (Grant No. 24C0316), the National Natural Science Foundation of China (Grant No. 12374046), Youth Science and Technology Talent Growth Project of the Education Department of Guizhou Province (project No. Qian Jiao Ji [2024]178), Shanghai Science and Technology Innovation Action Plan (Grant No. 24LZ1400800).